\documentstyle[12pt,axodraw]{article}
\begin{document}
\newcounter{myfn}[page]
\renewcommand{\thefootnote}{\fnsymbol{footnote}}
\newcommand{\myfootnote}[1]{\setcounter{footnote}{\value{footnote}}%
\footnote{#1}\stepcounter{footnote}}
\renewcommand{\theequation}{\thesection.\arabic{equation}}
\newcounter{saveeqn}
\newcommand{\add}{\addtocounter{equation}{1}}
\newcommand{\be}{\begin{equation}}
\newcommand{\ee}{\end{equation}}
\newcommand{\alpheqn}{\setcounter{saveeqn}{\value{equation}}%
\setcounter{equation}{0}%
\renewcommand{\theequation}{\mbox{\thesection.\arabic{saveeqn}{\alph{equation}}}}}
\newcommand{\reseteqn}{\setcounter{equation}{\value{saveeqn}}%
\renewcommand{\theequation}{\thesection.\arabic{equation}}}
\newenvironment{nedalph}{\add\alpheqn\begin{eqnarray}}{\end{eqnarray}\reseteqn}
\newsavebox{\PSLASH}
\sbox{\PSLASH}{$p$\hspace{-1.8mm}/}
\newcommand{\PS}{\usebox{\PSLASH}}
\newsavebox{\ASLASH}
\sbox{\ASLASH}{$A$\hspace{-1.8mm}/}
\newcommand{\AS}{\usebox{\ASLASH}}
\newsavebox{\KSLASH}
\sbox{\KSLASH}{$k$\hspace{-1.8mm}/}
\newcommand{\KS}{\usebox{\KSLASH}}
\newsavebox{\LSLASH}
\sbox{\LSLASH}{$\ell$\hspace{-1.8mm}/}
\newcommand{\LS}{\usebox{\LSLASH}}
\newsavebox{\SSLASH}
\sbox{\SSLASH}{$s$\hspace{-1.8mm}/}
\newcommand{\SS}{\usebox{\SSLASH}}
\newsavebox{\DSLASH}
\sbox{\DSLASH}{$D$\hspace{-2.4mm}/}
\newcommand{\DS}{\usebox{\DSLASH}}
\begin{center}
{\LARGE\bf{Phenomenological Consequences of Non-commutative QED }}\\
\vspace{1.5cm}
{\bf H.Arfaei} \footnote{\normalsize{Electronic address: Arfaei@theory.ipm.ac.ir}} and \hspace{1mm} {\bf M.H.Yavartanoo}  \footnote{\normalsize{Electronic address: Yavar@theory.ipm.ac.ir}}  \\
\vspace{1cm}

{\sl  Department of Physics, Sharif University of Technology}\\
{\sl P.O. Box 11365-9161, Tehran-Iran}\\and\\
{\sl{  Institute for Studies in Theoretical Physics and Mathematics (IPM)}}\\
{\sl{P.O. Box 19395-5531, Tehran-Iran}}\\
 \end{center}
  \vspace{2cm}
\vspace{0.5cm} \hspace{6.5cm} {\bf{Abstract}}
 \\
 In the context of the noncommutative
QED we consider few phenomena which reflect the noncommutativity.
In all of them  the new interactions  in the Feynmann diagrams
that are responsible for the deviation from the standard QED
results. These deviations appear as the violations of Lorentz
symmetry. We suggest experimental situations where these effects
may be observed. The extra phases have far reaching consequences
including violation of crossing symmetry. Considering the
$e\hspace{1mm}p$ scattering and Compton scattering the electric
dipole moments of the electron and the photon is calculated.
\par\noindent
\vskip1cm
\newpage
\section{Introduction}
Noncommutative gauge theories has been the center of extensive
attention[1-11].This interest was generated after the work of CDS
\cite{cds} and acquired boost after its derivation from string
theory \cite{NCGT}. The main property of the noncommutative gauge
theory is the existence of a non-zero constant background gauge
field induced from the nonzero antisymmetric form in the ten
dimensional space where the closed strings live. In the decoupling
limit where the open strings propagate only on the branes the
effective theory is a noncommutative gauge theory or
noncommutative open string theory  where the parameters
characterizing noncommutativity is derived from the background
field \cite{NCGT} \cite{ncft} \cite{NCQED}. There are two
essential points that such theories deviate from the standard
gauge theories. One is  the breakdown of Lorentz invariance, since
obviously a non-zero gauge field strength shows preferred
directions, and the other is the introduction of new interaction
(three photon vertex ) and modification of the standard ones. This
two properties have common origin and will obviously  appear in a
number of phenomena. In this article we shall consider certain
scattering processes where the Lorentz symmetry breaking will
manifestly appear . We will choose the simplest set up, the U(1)
theory and fermions coupled to it, i.e. noncommutative
QED\cite{NCQED} \cite{anomaly1}. We follow the notation of
\cite{anomaly1} where the Feynmann rules are also derived. We
consider four different phenomena; $ e e , e^+e^-$  scattering,
correction to the electrons magnetic moment, and the Compton
scattering .
 An interesting result is that the extended nature of the strings is
 revealed by observing the dipole dipole interaction of electrons.
 Scattering of photon on electron shows that they also carry electric
 dipole proportional to its momentum and square of the
 noncommutativity parameter.
In the following we shall briefly review the Feynman rules and
establish our notation. Section 2 presents the results of the
application of noncommutative QED to $ e e $and $e^+ e^- $
scattering . In section 3, we study electron proton interaction
and  derive the low energy amplitude from which the effective
potential is derived. It shows a correction to the Coulomb
potential which is interpreted as a velocity dependent dipole
moment of the electron. This velocity dependent dipole can be
understood in the frame work of string theory where the
non-trivial background stretches the moving string. Section four
is devoted to the Compton scattering . In this case the exchange
of photon results in a t-channel contribution to the amplitude
which is a characteristic of noncommutative QED. In this case we
observe a momentum dependent dipole moment for the photon which is
twice as large as that of the electron. \par As an example of
radiative correction we consider the anamolous magnetic moment of
electron. We see that $\overrightarrow{\mu}$ acquires a spin
independent term proportional to noncommutative parameter. The
conclusion is devoted to discussion on possible experiments that
may test the Lorentz symmetry breaking and the new interactions.\\
We start with the noncommutative action for electromagnetic field,
\begin{eqnarray}
S_{G}= -\frac{1}{4}\int d^{D}x \ F_{\mu\nu}\left(x\right)\star
F^{\mu\nu}\left(x\right).
\end{eqnarray}
\\
  where$F_{\mu\nu}$ define by
 \begin{eqnarray}\label{F27}
F_{\mu\nu}\left(x\right)&\equiv&
\partial_{\mu}A_{\nu}\left(x\right)-\partial_{\nu}A_{\mu}\left(x\right)+ig\big[A_{\mu}\left(x\right),A_{\nu}\left(x\right)\big]_{\star}.
\end{eqnarray}
The noncommutativity is
coded in the star product given by
 \\
 \begin{eqnarray} f\left(x\right)\star g\left(x\right)\equiv
e^{\frac{i\theta_{\mu\nu}}{2}\ \frac{\partial}{\partial\xi_{\mu}}\
\frac{\partial}{\partial\zeta_{\nu}}
}f\left(x+\xi\right)g\left(x+\zeta\right)\bigg|_{\xi=\zeta=0},
\end{eqnarray}
\begin{eqnarray}\label{F22}
[x_{\mu},x_{\nu}]=i\theta_{\mu\nu},
\end{eqnarray}
The fermion fields are  introduced by the action \cite{anomaly2} :
\begin{eqnarray}\label{F214}
S_{F}[\overline{\psi},\psi]= \int d^{D}x
\bigg[i\overline{\psi}\left(x\right)\gamma^{\mu}\star
D_{\mu}\psi\left(x\right)-m\overline{\psi}\left(x\right)\star\psi\left(x\right)\bigg],
\end{eqnarray}
where the covariant derivative is defined by:
\begin{eqnarray}\label{F215}
D_{\mu}\psi\left(x\right)\equiv\partial_{\mu}\psi\left(x\right)+igA_{\mu}\left(x\right)\star\psi\left(x\right).
\end{eqnarray}
For completeness we
quote the Feynman rules. \\
Fermion propagator :
\begin{nedalph}\label{FX1a}
\SetScale{0.8}
    \begin{picture}(80,20)(0,0)
    \Vertex(-170,0){2}
    \Line(-170,0)(-100,0)
    \Vertex(-100,0){2}
    \Text(-105,-10)[]{$p$}
    \end{picture}
  \frac{i}{\PS-m},
\nonumber
\end{eqnarray}
\\
Photon propagator :
\begin{eqnarray}\label{FX1b}
\SetScale{0.8}
    \begin{picture}(80,20)(0,0)
    \Vertex(-170,0){2}
    \Photon(-170,0)(-100,0){3}{7}
    \Vertex(-100,0){2}
    \Text(-108,10)[]{$k$}
    \Text(-135,-10)[]{$\mu$}
    \Text(-80,-10)[]{$\nu$}
    \end{picture}
  \frac{-ig_{\mu\nu}}{k^{2}}.\nonumber
\end{eqnarray}
\\
Fermion photon vertex : \vskip0.2cm
 \begin{eqnarray}\label{FX1c}
   \SetScale{.8}
    \begin{picture}(50,20)(0,0)
    \Vertex(-100,0){2}
    \Photon(-100,0)(-100,20){2}{4}
    \LongArrow(-90,18)(-90,12)
    \ArrowLine(-120,-20)(-100,0)
    \ArrowLine(-80,-20)(-100,0)
    \Text(-80,30)[]{$k_{1},\mu$}
    \Text(-110,-15)[]{$p_{1}$}
    \Text(-55,-15)[]{$p_{2}$}
    \end{picture}
 ie\gamma_{\mu}\ \mbox{exp}\left(- \frac{i}{2} p_{1} \times
p_{2}\right)\\
\nonumber
\end{eqnarray}
Three photon vertex : \vskip0.2cm
\begin{eqnarray}\label{FX1d}
   \SetScale{0.8}
    \begin{picture}(600,35)(0,0)
    \Vertex(75,20){2}
    \Photon(75,20)(75,40){2}{4}
    \LongArrow(65,40)(65,32)
    \LongArrow(50,5)(55,12)
    \LongArrow(100,5)(95,12)
    \Photon(55,0)(75,20){2}{4}
    \Photon(95,0)(75,20){2}{4}
    \Text(60,45)[]{$k_{1},\mu_{1}$}
    \Text(15,5)[]{$k_{2},\mu_{2}$}
    \Text(100,5)[]{$k_{3},\mu_{3}$}
    \Text(175,40)[]{$-2e \sin\left(\frac{i}{2}k_{1} \times k_{2}\right)$}
    \Text(292,15)[]{$\times \left[g_{\mu_{1}\mu_{2}}\left(k_{1}-k_{2}\right)_{\mu_{3}} +
                                  g_{\mu_{1}\mu_{3}}\left(k_{3}-k_{1}\right)_{\mu_{2}}+
                                  g_{\mu_{3}\mu_{2}}\left(k_{2}-k_{3}\right)_{\mu_{1}} \right]$}
\end{picture}
\nonumber
\end{nedalph}

Four photon vertex :
\begin{eqnarray}
\begin{picture}(600,45)(0,0)
    \Vertex(60,0){2}
    \Photon(80,20)(60,0){2}{4}
    \Photon(40,20)(60,0){2}{4}
    \LongArrow(35,-15)(40,-8)
    \LongArrow(86,-15)(80,-8)
    \LongArrow(86,15)(80,8)
    \LongArrow(35,15)(40,8)
    \Photon(40,-20)(60,0){2}{4}
    \Photon(80,-20)(60,0){2}{4}
    \Text(26,-25)[]{$k_{2},\mu_{2}$}
    \Text(93,-25)[]{$k_{3},\mu_{3}$}
    \Text(93,25)[]{$k_{4},\mu_{4}$}
    \Text(26,25)[]{$k_{1},\mu_{1}$}
    \Text(140,20)[]{$\-4ie^2 $}
    \Text(161,20)[]{\Huge [  }
\Text(302,20)[]{$(g_{\mu_{1}\mu_{3}}g_{\mu_{2}\mu_{4}}-g_{\mu_{1}\mu_{4}}
g_{\mu_{2}\mu_{3}}) \sin\left(\frac{1}{2} k_{1}\times k_{2}
\right)sin\left(\frac{1}{2} k_{3}\times k_{4} \right)$}
\Text(302,0)[]{$+(g_{\mu_{1}\mu_{4}}g_{\mu_{2}\mu_{3}}-g_{\mu_{1}\mu_{2}}
g_{\mu_{3}\mu_{4}})\sin\left(\frac{1}{2} k_{3}\times k_{1}
\right)sin\left(\frac{1}{2} k_{2}\times k_{4} \right)$}
\Text(307,-20)[]{$+(g_{\mu_{1}\mu_{2}}g_{\mu_{3}\mu_{4}}-g_{\mu_{1}\mu_{3}}
g_{\mu_{2}\mu_{4}})\sin\left(\frac{1}{2} k_{1}\times k_{4}
\right)sin\left(\frac{1}{2} k_{2}\times k_{3} \right)$}
\Text(455,-20)[]{\Huge ]}
    \end{picture}
\end{eqnarray}
\\
\\
\\
In above expression we have , $p \times q :=
p^{\mu}\theta_{\mu\nu}q^{\nu}$. Now we consider standard processes
but in the new noncommutative context.
\section{$e \hspace{3mm} e$ and $e^{+} \hspace{1mm} e^{-} $ scattering}
\par
As our first example consider $e e \rightarrow e e $. The
following diagrams take part in the $e \hspace{1mm} e$
scattering\\
\vspace{12mm}
\begin{eqnarray}
\SetScale{0.8}
    \begin{picture}(600,80)(0,0)
     \Vertex(20,100){1}
     \Vertex(50,100){1}
     \ArrowLine(0,50)(20,100)
     \Photon(20,100)(50,100){2}{7}
     \ArrowLine(20,100)(0,150)
     \ArrowLine(70,50)(50,100)
     \ArrowLine(50,100)(70,150)
     \Text(-8,45)[]{$p_1$}
     \Text(-8,125)[]{$p_4$}
     \Text(68,125)[]{$p_3$}
     \Text(68,45)[]{$p_2$}
      \Text(80,80)[1]{$ {\cal{M}}_{t}= -$}
     \Text(110,80)[1]{\LARGE$\frac{e^2}{t}$}
     \Text(270,80)[1]{$ \left(\overline{U}(p_4)\gamma^{\mu} U(p_1)\right)\left(\overline{U}(p_3)\gamma_{\mu} U(p_2)\right)\exp{\frac{i}{2}(p_4 \times p_1 + p_2\times p_3)}$}
       \end{picture}
    \end{eqnarray}
\begin{eqnarray}
\SetScale{0.8}
    \begin{picture}(600,80)(0,0)
     \Vertex(20,100){1}
     \Vertex(50,100){1}
     \ArrowLine(0,50)(20,100)
     \Photon(20,100)(50,100){2}{7}
     \ArrowLine(20,100)(70,160)
     \ArrowLine(70,50)(50,100)
     \ArrowLine(50,100)(0,160)
     \Text(-8,45)[]{$p_1$}
     \Text(-8,125)[]{$p_4$}
     \Text(68,125)[]{$p_3$}
     \Text(68,45)[]{$p_2$}
     \Text(80,80)[1]{$ {\cal{M}}_{u}= -$}
     \Text(110,80)[1]{\LARGE$\frac{e^2}{u}$}
     \Text(270,80)[1]{$ \left(\overline{U}(p_3)\gamma^{\mu} U(p_1)\right)\left(\overline{U}(p_4)\gamma_{\mu} U(p_2)\right)\exp{\frac{i}{2}(p_3 \times p_1 + p_2\times p_4)}$}
    \end{picture}
    \end{eqnarray}
$p_1, p_2, p_3$ and $p_4$ are the momenta of the electorns as show
in the above diagrams.\\
The only new factors are the phases which changes the interference
terms in the cross section.
 Then in high energy region partial cross sections is proportional to: \\
\\
 \be{ \cal{|M|}}^2=2 e^2 \left( \frac{s^2+u^2}{t^2}
+\frac{s^2+t^2}{u^2} + \frac{2s^2}{ut} cos(p_3 \times p_4)\right)
\ee As one can see the interference term makes a difference when
the outgoing momenta are not along the total incoming momentum
$P=p_1+p_2$. The origin of this interference is that in the two
Feynman diagrams the outgoing electrons are relatively crossed and
their phases are different. Unfortunately this difference vanishes
in the center of mass frame. Therefore to set up an experiment to
observe the interference one can devise a situation where  the two
beams are of different energy or are not colinear.
\\
If $\Theta_i :=\epsilon_{ijk}\theta_{jk}$ and $
\overrightarrow{p}_{r \bot} $ is the projection of momentum $
\overrightarrow{p}_r $ on the plane perpendicular to
$\overrightarrow{\Theta}$ , \be \cos(p_3 \times p_4)=cos(p_3
\times p ) = \cos( \theta p_3 p_{\bot} \sin\alpha ) \ee
 where$ \overrightarrow{p}=p_1+p_2$ and $ \sin \alpha $ is angle
 between $\overrightarrow{p}_{\bot}$ and
 $\overrightarrow{p}_{\bot}^3$.
 Hence the cross section goes under periodic change when the angle
 of the perpendicular component of the outgoing electron changes.
 If ${\cal{M}}_0$ is the amplitude in the standard QED then
 \be
 \Delta |{\cal{M}}| ^2 = |{\cal{M}}_0 |^2 -|{\cal{M}}| ^2
 =\frac{2s^2}{ut}\left(cos\theta \hspace{1mm}p_{\bot}^3 p_{\bot}
 sin\alpha \right) .
 \ee
 One can see that the partial cross section decrease in the
 noncommutative case.
 In the forward direction the cross section is unchanged.
 An interesting and important point is that the change in
 noncommutative cross section depends on the frame of reference.
 In the center of mass frame
 $\overrightarrow{p}=\overrightarrow{p}_1 + \overrightarrow{p}_2 =
 0$
 and hence $ \Delta |{\cal{M}} ^2 |=0$ so we do not observe any
 effect. In the laboratory frame when the target is stationary
 it is  essentially observable.  The dependence on the reference  frame is a reflection on
 breaking of Lorentz invariance group $SO(3,1) $ to $SO(2) \times
 SO(1,1)$.
 The  $e^{+} \hspace{1mm} e^{-} $  scattering shows no such
interference term as can be seen in the following expressions for
the amplitude. \be { |\cal{M}}|^2=2 e^2 \left( \frac{s^2+u^2}{t^2}
+\frac{s^2+t^2}{u^2} + \frac{2s^2}{ut} \right) \ee The difference
appears because in this case we have only planner diagrams where
the phases are the same. This difference is an example of the
violation of simple crossing symmetry! Crossing symmetry break
down can be understood since s, t, and u are no longer the
invariant parameters, the Lorentz symmetry is broken.
\\
\section{electron proton scattering}
In this part we consider the scattering of two fermions, one heavy
and one light which we refer to  as proton and electron. The main
object from which we may draw how energy observable is the
scattering amplitude. To see the effect  of noncommutativity the
tree diagram is sufficient in contrast to the problem of anomalous
magnetic moment where we have to consider the one loop diagrams.
Using the Feynman rules given in previous section, the tree
amplitude picks up  a phase with respect to the commutative case
giving.
 \vspace{45mm}
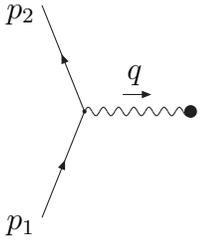
\begin{eqnarray}
\SetScale{0.8}
    \begin{picture}(600,40)(0,0)
     \Vertex(20,80){1}
     \Vertex(70,80){3}
     \ArrowLine(0,30)(20,80)
     \Photon(20,80)(70,80){2}{7}
     \ArrowLine(20,80)(0,130)
     \LongArrow(38,88)(50,88)
     \Text(35,78)[]{$q$}
     \Text(-8,21)[]{$p_1$}
     \Text(-8,101)[1]{$p_2$}
     \Text(123,96)[1]{$\cal{M_{NC}}$}
     \Text(170,96)[1]{$= e^{\frac{-i}{2} P_1\times p_2} $}
    \Text(229,96)[1]{$\cal{M_{C}} ,$}
     \Text(125,56)[1]{where}
      \Text(123,20)[1]{$\cal{M_{C}}$}
       \Text(170,20)[1]{ $ = i \hspace{1mm} e \hspace{1mm}\gamma^{\mu} \hspace{1mm}\frac{g_{\mu\nu}}{q^2}$}
       \Text(217,20)[1]{\Large $j^{\nu}$}
      \end{picture}
       \end{eqnarray}
where the kinematical parameters are shown on the Feynman diagram.
$ j^{\nu} $ is the external current. In the limit where
$q\rightarrow 0$ this amplitude gives the potential of an
stationary charged particle which deviates from the Coulomb
potential, this deviation is a function of $ $ the momentum of the
lighter particle. \be V=e^2\int \frac{d^{3}q}{(2\pi)^3} e^{i
q.(r+\frac{\widetilde{P}}{2})} \hspace{2.5mm} = \hspace{2.5mm}
\frac{e^2}{4\pi} \frac{1}{|r+\frac{\widetilde{P}}{2}|} \ee This
potential is very different from the usual concept. It is
hamiltonian between electron and proton in the sense that the
amplitude calculated in Born approximation yields the tree
amplitude.The shift of the singularity from $r=0$ to
$r=-\frac{\widetilde{p}}{2}$ can be explained by taking the string
theory point of view. Open string moving with the ends on the
brane are stretched, the amount of which is proportional to its
momentum \cite{string}. The direction of the stretch is
perpendicular to $\overrightarrow{\Theta}$ and
$\overrightarrow{p}$. So effectively the electric charge is moved
$ \frac{1}{2} \overrightarrow{p}\times \overrightarrow{\Theta}$
with respect to the center of mass of the string. Obviously such
potential violates rotational invariance . The background $
\theta^{ij}$ specifies a particular direction in space given by
vector $\Theta_i=\epsilon_{ijk} \theta^{jk}$. We still have
rotational invariance around $\overrightarrow{\Theta}$. For a
given value of momentum the interaction strongly depends on the
direction of the momentum of the projectile. If it moves parallel
to the vector $\overrightarrow{\Theta} = p_i\theta^{ij} $ it sees
only the standard Coulomb force because $\widetilde{p}$ which sets
the strength isof the dipole is zero. The extra term is maximum
when the momentum is in the plane perpendicular to the
$\overrightarrow{\Theta}$. If we expand this potential we have :
\be V=\frac{e^2}{4\pi}\left[ \frac{1}{|r|}
-\frac{\overrightarrow{r}.\widetilde{p}}{2r^3}\right]\ee which can
be interpreted to the projectile having dipole moment proportional
to its momentum. Let us naively calculate the force due to this
term,i.e dipole force \be \textit{F}=\frac{e^2}{8\pi
r^5}\left[r^2\theta_{ij} p_j-3r_k\theta_{kj}p_jr_i \right] \ee The
first term looks like a magnetic field acting on the electron. But
the second term gives a central force proportional to the momentum
perpendicular to the background field $ \overrightarrow{\Theta} $.
Therefore the signature of noncommutativity in $e \hspace{1mm} p$
scattering is velocity dependent radial force! in which varies as
the inverse third power of the distance between the two particles.
Of course the potential which is useful and reliable for long
separation shows contribution from higher modes. Certainly short
distance behavior of the potential will be affected by correction
to the propagator  from photon and electron loops which we defer
to further investigation. Such contributions are of orders
$\theta^2 $ which must be very small.
\section{Compton Scattering}
Another phenomenon that may reveal the non commutativity of space
is the electron photon (Compton) scattering.\\
We assume that the charged particle e.g electron is massive. The
noncommutative theories predict direct interaction of three
$\textit{abelian gauge particle}$ which give rises to a t-channel
amplitude to the Compton scattering via the following Feynman
diagram :
\\
\\
\vspace{4.5cm}
\begin{eqnarray}
\SetScale{0.8}
    \begin{picture}(600,80)(0,0)
     \Vertex(20,100){1}
     \Vertex(50,100){1}
     \ArrowLine(0,50)(20,100)
     \Photon(20,100)(50,100){2}{7}
     \ArrowLine(20,100)(0,150)
     \Photon(70,50)(50,100){2}{7}
     \Photon(50,100)(70,150){2}{7}
     \LongArrow(78,150)(82,158)
     \LongArrow(78,55)(74,63)
     \Text(-8,45)[]{$p_1$}
     \Text(-8,125)[]{$p_2$}
     \Text(75,125)[]{$k_2$}
     \Text(73,45)[]{$k_1$}
     \Text(240,100)[1]{${\cal{M}}_t
=-2 i \frac{e^2}{t}\epsilon_{\mu}(k_1)\epsilon^{\ast}_{\nu}(k_2)
     sin(\frac{k_2 \times k_1}{2})   \exp{\frac{i}{2}(p_1\times p_2)} \overline{U}(p_2) \gamma_{\rho} U(p_1)
     C^{\mu\rho\nu} ,
$}
      \Text(100,70)[]{where}
      \Text(250,40)[]{$C^{\mu\rho\nu} =
      (k_2-2k_1)^{\nu} g^{\rho\mu}+(k_1-2k_2)^{\mu} g^{\rho\nu}
+(k_1+k_2)^{rho}g^{\mu\nu} .$}
       \end{picture}
    \end{eqnarray}
\vspace{0cm} This diagram has contribution to the amplitude where
its leading share to the cross section come from an interference
with the standard term which are represented in the following
diagram : \\      \\
\begin{eqnarray}
\SetScale{0.8}
    \begin{picture}(600,80)(0,0)
     \Vertex(20,80){1}
     \Vertex(20,30){1}
     \ArrowLine(0,0)(20,30)
     \Photon(40,0)(20,30){2}{7}
     \ArrowLine(20,80)(0,110)
     \Photon(40,110)(20,80){2}{7}
     \ArrowLine(20,30)(20,80)
     \Text(-5,-5)[]{$p_1$}
     \Text(43,-5)[]{$k_1$}
     \Text(-5,95)[]{$p_2$}
     \Text(43,95)[]{$k_2$}
     \LongArrow(35,113)(40,121)
     \LongArrow(48,5)(44,13)
     \Text(250,60)[1]{$
     {\cal{M}}_s=${\Large{$\frac{e^2}{s-m^2}$}}$\epsilon_{\alpha}(k_1)\epsilon^{\ast}_{\beta}(k_2)
     \overline{U}(p_2) \gamma^{\beta} \left(\PS_1 +\KS_1+m\right)\gamma^{\alpha}U(p_1)$}
     \Text(245,30)[1]{$\exp{\frac{i}{2}(p_1 \times k_1 + p_1\times p_2+ k_1 \times p_2)}$}
    \end{picture}
    \end{eqnarray}
\\   \\
\begin{eqnarray}
\SetScale{0.8}
    \begin{picture}(600,80)(0,0)
     \Vertex(20,80){1}
     \Vertex(20,30){1}
     \ArrowLine(0,0)(20,30)
     \Photon(45,110)(20,30){2}{8}
     \ArrowLine(20,80)(0,110)
     \Photon(45,0)(20,80){2}{8}
     \ArrowLine(20,30)(20,80)
     \Text(-5,0)[]{$p_1$}
     \Text(50,0)[]{$k_1$}
     \Text(-5,90)[]{$p_2$}
     \Text(50,95)[]{$k_2$}
     \LongArrow(48,105)(53,113)
     \LongArrow(53,5)(49,13)
     \Text(250,60)[1]{$
     {\cal{M}}_u= $ {\Large{$
   \frac{e^2}{u-m^2}$}}$ \epsilon_{\beta}(k_1)\epsilon^{\ast}_{\alpha}(k_2)
     \overline{U}(p_2) \gamma^{\beta} \left(\PS_1
     -\KS_2+m\right)\gamma^{\alpha}U(p_1)$}
     \Text(245,30)[1]{$\exp{\frac{i}{2}(p_1 \times k_1 + p_1\times p_2+ k_1 \times p_2)} $}
    \end{picture}
    \end{eqnarray}
the amplitude cross section is proportional
  to   \be |{\cal{M}}_{NC}|^2 =
|{\cal{M}}_{C}|^2 + {\large{\delta}}|{\cal{M}}|^2 ,\ee
 which $ |{\cal{M}}_{C}|^2 $ term is same as commutative case, and
 ${\large{\delta}}|{\cal{M}}|^2$ is additional contribution in NCQED.
 \begin{eqnarray}
{\large{\delta}}|{\cal{M}}|^2 = 4 e^4 sin^2(\frac{k_1 \times
k_2}{2})\left( \frac{s^2+2st-3m^2s-2m^2t+4m^4}{st} -
\frac{u^2+2ut-3m^2u-2m^2t+4m^4}{ut} \right) \nonumber
\end{eqnarray}
       \vspace{-10mm}
\begin{eqnarray}
\hspace{-45mm} + 16e^4 sin^2(\frac{k_1 \times
k_2}{2})\left(1-\frac{5us}{4}-\frac{5m^2}{4t}
+\frac{5m^4}{4t^2}\right) \nonumber
 \end{eqnarray}
   \vspace{-10mm}
 \begin{eqnarray}
 \hspace{-50mm} +\frac{4 e^4}{us} \left(1-cos(k_1 \times
 k_2)\right) \left( m^2(s+u)+2m^4 \right)
 \end{eqnarray}
 In high energy limit ,the leading term is
  \be
 e^4(k_1 \times k_2 )^2 \left( \frac{s}{t} + \frac{u}{t}
 -\frac{4}{t^2}
 -\frac{5us}{t^2}\right)
 \ee
although classically it is not appropriate to use the concept of
potential between proton and electron, we consider the Fourier
transformation  of the partial amplitude as an effective potential
that may shed light on the electromagnetic interaction of photon
and electron. As in the case of fermions we take the situation
where the polarization is unchanged. It turns out   that such
effective potential is \be V=\frac{2e^2k_0
\cos(\alpha)}{4\pi}\left( \frac{1}{|r+\frac{\widetilde{P}}{2}|}
-\frac{1}{|r-\frac{\widetilde{P}}{2}|}\right)\ee
 this clearly
shows that the photon is effectively seen by the electron as two
separated charges with values $\pm e$  and separation
$\widetilde{P}$. In the large $r$ approximation that we are
considering it is natural to expand the potential in powers of
$\frac{1}{r}$ to final the dipole moment of the photon to be : \be
e \widetilde{P} = e \theta_{ij} p_j = e \epsilon_{ijk}
\overrightarrow{\Theta_k}p_j = e \overrightarrow{P} \times
\overrightarrow{\Theta}. \ee
 Note that P in the photon momentum.
It is interesting that the photons electric dipole moment is twice
as big as that of an electron for the same momentum P. Again we
may use the string theory point of view to understand the two
terms in above formula. The two ends of the photon are opposite
charges separated from each other by $\widetilde{p} $.
 This separation gives a dipole moment $\overrightarrow{p} \times \overrightarrow{\Theta} . $
 To identify a signature of noncommutativity in Compton
scattering we look at the cross section. Breakdown of rotational
invariance in obvious from the contribution of $k$ to  the
amplitude ( or equivalently the effective potential ). If the
photon   beam is rotated we see a change in the Compton cross
section. The interference is maximum when p is perpendicular to
$\overrightarrow{\Theta}$ and goes to zero when p is parallel to
it. Therefore a change in the Compton cross section with rotation
of the photon beam is a possible signature of the space-space
noncommutativity. the variation with orientation of the photon
beam is independent of polarization when the $ \cos(\alpha) $ in $
(4.25)$ is averaged out.
\section{Anomalous magnetic momentum of electron }
The effects we have discussed  in previous sections are all from
leading tree diagrams. In this section we consider a one loop
effect. i.e electron's anomalous magnetic moment.\footnote{when
our work on this part was completed the paper [0009037] appear
with the same ? section IV of our work}
\\ We will find that electron's magnetic moment dose
not receive any correction from noncommutativity of space. The
correction to the vertex is relativistic effect. To the lowest
order the contribution come from \vspace{3cm}
\begin{eqnarray}
\SetScale{0.8}
    \begin{picture}(600,50)(0,0)
     \Vertex(40,100){1}
     \Vertex(90,100){3}
     \ArrowLine(20,50)(40,100)
     \Photon(40,100)(90,100){2}{7}
     \ArrowLine(40,100)(20,150)
     \LongArrow(58,110)(70,110)
     \Text(55,97)[]{$q$}
     \Text(12,42)[]{$p_1$}
     \Text(12,125)[]{$p_2$}
     \Text(180,100)[]{ whose value is proportional to}
    \Text(150,40)[]{ $ e^{\frac{i}{2} p' \times p }\overline{U}(p)\gamma_{\mu}U(p')$}
     \end{picture}
       \end{eqnarray}
\vspace{-5mm}
 the correction to the vertex comes from the diagrams, \\
 \begin{eqnarray}
\SetScale{0.8}
    \begin{picture}(1500,150)(0,0)
     \Vertex(360,100){1}              \Vertex(60,100){1}
     \Vertex(410,100){3}              \Vertex(110,100){3}
     \ArrowLine(360,100)(328,60)     \ArrowLine(28,60)(36,70)
     \ArrowLine(360,100)(328,140)      \ArrowLine(36,70)(36,130)
     \Photon(336,70)(336,130){2}{7}  \Photon(60,100)(36,70){2}{5}
     \Photon(360,100)(410,100){2}{7}  \Photon(60,100)(110,100){2}{7}
     \LongArrow(385,110)(373,110)     \Photon(60,100)(36,130){2}{5}
     \Text(305,97)[]{$q$}             \ArrowLine(36,130)(28,140)
     \Text(253,45)[]{$p$}              \LongArrow(85,110)(73,110)
     \Text(253,115)[]{$p'$}            \Text(65,97)[]{$q$}
     \Text(310,15)[]{$(a)$}           \Text(12,45)[]{$p$}
                                      \Text(12,115)[]{$p'$}
                                       \Text(60,15)[]{$(b)$}
                                       \end{picture}
                                       \end{eqnarray}
\\
where the corresponding contributions to the vertex function are :
\newpage
\begin{eqnarray} i e \Gamma^{\mu}_a e^{(\frac{i}{2} p' \times  p)} =
e^{(\frac{i}{2} p' \times p)} \int \frac{d^4k}{(2\pi)^4}
\overline{U}(p') \left[ i e \gamma^{\alpha} \frac{i(\PS'-\KS +m)}
{(p'-k)^2-m^2} i e \gamma^{\mu} \frac{i(\PS-\KS +m)}
{(p-k)^2-m^2}i e\gamma^{\beta} \frac{-ig_{\alpha\beta}}{k^2}
\right] \overline{U}(p) \nonumber \end{eqnarray}
\begin{eqnarray} i e \Gamma^{\mu}_b e^{(\frac{i}{2} p' \times  p)} =
e^{(\frac{i}{2} p' \times p)} \int \frac{d^4k}{(2\pi)^4}(1-e^{(ik
\times q)}) \overline{U}(p') \left[ i e \gamma^{\alpha}
\frac{i(\PS-\KS +m)} {(p'-k)^2-m^2} i e \gamma^{\beta}
\frac{-ig_{\beta\nu}}{k^2}\frac{-ig_{\alpha\beta}}{(k+q)^2}
\right]  \nonumber
\end{eqnarray}
\vspace{-10mm}
\begin{eqnarray}
\hspace{4.5cm}  i e C^{\nu\mu\rho}U(p)
\end{eqnarray}
which $C^{\nu\mu\rho} = (k-q)^{\rho} g^{\mu\nu} (2q+k)^{\nu}
g^{\mu\rho} -(2k+q)^{\mu} g^{\rho\nu} .$ \\ The diagram 'a' except
for an over all phase is the same as in the commutative case. Its
phase vanishes when $q\rightarrow 0$. Diagram 'b' dose not appear
in commutative QED and add a new contribution to electron magnetic
dipole. \be <\overrightarrow{\mu}>=\frac{e}{2m} (
2+\frac{\alpha}{2\pi})\overrightarrow{S} +\frac{e \alpha \gamma_E
m}{6\pi} \overrightarrow{\theta} \ee This extra term is a constant
independent of electron's kinematical  state, sipn or momentum. To
see such independent magnetic moment we may use a Stern-Gerlach
apparatus. The two electron beams of two different spin respond
differently to the external non-uniform magnetic field. The two
eigen state of spin have the magnetic moments $
\mu_{\pm}=\pm\frac{e}{4m}( 2+\frac{\alpha}{2\pi})\hbar +\frac{e
\alpha \gamma_E m}{6\pi} \theta$. It is worth noting that spin
resonance experiments can not reveal the noncommutativity because
they are sensitive to the energy difference of the two states.
This difference is independent of $\theta$.
\section{Conclusion}
\par
We have considered a number of processes  in which the
noncommutativity manifests itself . This manifestation is through
breaking of Lorentz invariance (both rotation and boost). This
symmetry breaking  cannot be observed in the scattering processes
unless we are in a frame different from the center of mass frame
and the momenta are not colinear. As we discussed in the
scattering of $(e e, e^{+}e^{-}, \gamma e,e p, )$ the signature
for noncommutativity is the change of the partial and total cross
section  with rotation of the incident beams or out going detected
particle. This change occurs mainly when the scattered electron
makes a large angle, close to $\pi /2$ with  the direction of the
incident beam. The characteristic of this change is decrease in
the cross section which may distinguish it from other sources of
anisotropy such as dipole effect due to motion with respect to the
microwave background radiation. The other difference with such
anisotropy is that the noncommutativity may specify a direction
which is different from the detected motion of earth in the cosmic
thermal background. The change in the cross section is very small,
of the order of $(\Theta^2)$ which is not easy to measure. One may
set a bound for it by relating it to other physical quantities
such as axion expectation value. In all the scattering cases
change in the direction of the beam may not be possible to perform
in the laboratory. An obvious suggestion is the use of earth's
rotation. Hence a comparison of the measurements of $e e$ or $e^ +
e^-$ and also Compton scattering partial cross sections in
different times of the day and year may indicate anisotropy due to
noncommutativity.\\
 In the case of electron anomalous magnetic
moment, although noncommutativity predicts a constant magnetic
moment independent of the spin , shift in the frequency of spin
resonance remains unchanged. This happens because the spin
dependent part of the anomalous magnetic moment turns out to be
the same as in the as the commutative case. This forces the spin
flip energy to remain unchanged. In this case an experiment like
that of Stern-Gerlach may be useful. The magnetic force on an
electron is proportional to its magnetic moment which are
different in magnitude for the two spin orientations. Therefore
asymmetry in the deviation of the up and down beams, after
correction for other effects is sensitive to noncommutativity.
\\Of course to higher order of perturbation more intricate
dependence on $\theta$ may be discovered which is of higher order
of $\alpha $ and $ \theta $. Such effects are theoretically
interesting and are under investigation by the authors but
certainly are far too small to be detected or sought for before
observation of leading effects.
\section{Acknowledgments}
We would like to thank F.Ardalan and N.Sadooghi for fruitful
discussions.

\end{document}